# Asymmetric four-grating compressor for ultrafast high power lasers

Xiong Shen[a#], Shuman Du[a,b#], Jun Liu[a,b,*], Ruxin Li[a,b]

[a] *State Key Laboratory of High Field Laser Physics and CAS Center for Excellence in Ultra-intense Laser Science, Shanghai Institute of Optics and Fine Mechanics, Chinese Academy of Sciences, Shanghai 201800, China.*

[b] *University Center of Materials Science and Optoelectronics Engineering, University of Chinese Academy of Sciences, Beijing 100049, China.*

[#] *These authors contributed equally to this work*

[*] *Corresponding author: jliu@siom.ac.cn*

**Abstract**

The peak power improvement and running safety of petawatt (PW) lasers are limited by laser-induced damage of optical components with limited sizes and damage thresholds. Diffraction gratings in pulse compressors have been the shortest stave of PW lasers up to now, as to manufacture a high quality meter-sized grating remains particularly challenging. Here, the asymmetric four-grating compressor (AFGC) with asymmetric configuration is proposed for PW lasers to increase the maximum bearable output pulse energy and running safety without neither additional optical component nor extra control in comparison to a traditional Treacy four-grating compressor (TFGC) with symmetric configuration. In AFGC, suitable spatial dispersion can be introduced in the output laser beam which is able to decrease the laser spatial intensity modulation (LSIM) of the output beam on the final grating. The introduced spatial dispersion can be automatically compensated at the focal plane by using the spatiotemporal focusing technique. Based on this simple AFGC design, not only the damage risk of the final grating can be reduced, but also the maximum output pulse energy can be improved by about 1.8 times theoretically. As an example, 100 PW output power can be achieved theoretically by using the AFGC with an input beam size of 550×700 mm$^2$.

**Introduction**

Petawatt (PW) lasers benefited from advanced optical techniques, especially the chirped pulse amplification (CPA) [1] and the optical parametric CPA (OPCPA) [2], are coming online across the world [3]. Since the demonstration of the first PW laser (Nova laser, 1.5 PW) in 1999, which produced 660 J energy in a compressed 440±20 fs pulse [4], PW class lasers have been swarmed into laboratories, and more than 50 facilities that are, or have been, operational, under construction, or in the planning/conceptual phase [3]. Among them, both the SULF facility in China (SULF-10 PW) and the ELI-Nuclear Physics (ELI-NP) in Romania have already achieved the highest peak power of 10 PW to date recently [5, 6].

Flourish researches in the relativistic optical science have been promoted by these PW class lasers that with focused intensities higher than 10$^{18}$ W/cm$^2$ [7]. The extremely high optical-field conditions provide powerful tools for many high-field researches, such as particle acceleration, high energy secondary sources generation, laboratory astrophysics, nonlinear QED and so on. Scientists are even imagining the use of a future 100 PW laser (SEL-100 PW), that has been started up in 2018

[8], in the frontier research of vacuum birefringence exploration with extreme $10^{23-24}$ W/cm² focused intensity [9]. These attractive applications have become a main motivation of PW lasers development in turn.

For PW lasers, the general principle is that, weak femtosecond seed pulses are stretched to nanosecond pulses and amplified to hundreds or even thousands of Joules in laser gain media by using the CPA or the OPCPA techniques, where the amplified nanosecond pulses are compressed back to femtosecond pulses by using diffraction grating based pulse compressors. Laser-induced damage of optical components is a main serious limitation in PW lasers. Optical components owning large sizes are usually used to reduce laser fluence. In PW lasers, diffraction gratings in pulse compressor have been the shortest stave because it is hard to manufacture a high quality meter-sized grating up to now, especially for 10s to 100s PW lasers [10, 11].

To overcome the diffraction grating limitation, tiled grating (or mosaic grating) method that a large-sized grating composed of several small-sized gratings was proposed for PW lasers [12]. It is easy to manufacture small-sized gratings, while the composed small-sized grating components must be tiled precisely with the six degrees of freedom for each other: piston and shift, groove spacing and tilt, and rotation and tip [11], which results in a new control challenge. Coherent beam combination (CBC) is another method that has been extensively studied for PW lasers [13]. Laser beams are split into $N$ sub-beams prior to amplifier [14], compressor [15, 16], or even in the compressor [11], and followed by recombining of the compressed sub-beams. However, the CBC method is complicated as it is sensitive to the differences of optical delay, pointing stability, wavefront, and dispersion among the sub-beams [11, 17]. Except for these two methods, a feasible multistep pulse compressor (MPC) method was proposed recently [18]. The method increases the bearable output pulse energy of the compressor by modifying beam spatiotemporal properties, which can lead a 100 PW output with single beam. To improve the bearable output pulse energy of PW lasers with available diffraction gratings up to now, is still worth of exploring no matter from the aspect of cost performance or complexity, which can also guarantee system opera ting in a safer zone below laser-induced damage threshold (LIDT) for the same energy output.

In this study, we propose an asymmetric four-grating compressor (AFGC) method that can increase the maximum bearable output pulse energy and running safety of ultrafast ultraintense laser systems. It is based on spatiotemporal property modification similar as the MPC method [18], while simply modified the widely used general four-grating compressor (FGC) into AFGC structure instead of prism pairs in pre-compressor. In this simple modification, there is no additional optical component needed, neither increasing the difficulty of compressor controlling process. The influences of the AFGC configuration to the output laser spatiotemporal properties are analyzed in detail. The laser beam output from the compressor owns spatial dispersion which can improve the maximum bearable output pulse energy and running safety of the FGC. Furthermore, the induced spatial dispersion can be automatically compensated at the focal plane by using spatiotemporal focusing technique, which induces a high peak power laser. By introducing a spatial dispersion of $d0$=60 mm in AFGC, the laser spatial intensity modulation (LSIM) can be decreased from 2 to 1.1. It means 1.8 times increasing of the output pulse energy from the AFGC is possible in comparison to a traditional Treacy four-grating compressor (TFGC) with symmetric configuration in theory. As a result, about 100 PW laser can be achieved theoretically with an input beam size of 550×700 mm².

**Limitation factors in four-grating compressor**

In a FGC, reflective diffraction gratings are always metal or multilayer dielectric coated with high diffraction efficiency. While gold-coated gratings are the choice of compressors for PW lasers even with hundreds of times low LIDT comparing to multilayer dielectric gratings [20]. This is because that gold-coated gratings have high diffraction efficiencies over a broadband spectra which can support tens femtoseconds pulses output. What's more, gold is also a noble metal can maintain its high property for a long time without the protect from a transparent oxide layer like silver or aluminum [10], and will suffer no delamination due to internal stress buildup in vacuum like protected silver or aluminum coated metallic gratings or multilayer dielectric gratings will suffer.

For gold-coated gratings, experimental data of LIDT laser fluence ($mJ/cm^2$) were concluded about 2-3 times higher for nanosecond pulse in comparison to femtosecond pulse [10, 18]. In a general TFGC, 1) the beam sizes on the first and final gratings are almost the same, which are smaller than that on the other two gratings, 2) pulses are compressed from several nanoseconds to tens of femtoseconds by the compressor in PW lasers, 3) the diffraction efficiencies can be guaranteed larger than 90% on each gratings. These three aspects lead to a laser fluence ratio on the first and the final gratings about F1:F4=1:$0.9^3$=1.37:1 (supposing the diffraction efficiency is 90% on each grating), which is obviously about two times smaller than their 2-3 times LIDT ratio. It can be concluded that although the first grating bears the highest pulse energy, while the final grating stand the most dangerous risk of laser-induced damage. In the condition of under LIDT of the final grating, the first grating can bear about twice of the incident laser pulse energy, which has been used in the in-house beam-splitting pulse compressor method for PW lasers proposed recently [11] for the maximum utilization of the first pair of gratings in PW laser compressors.

Besides the pulse duration affecting the LIDT of compression gratings, laser LSIM of the amplified beam is another important factor that affects the output pulse energy of the FGC. LSIM is generally defined as the ratio between the maximum local intensity to the average beam intensity, which is assumed to be about 2 for a general amplified beam of PW lasers. To avoid laser-induced damage of the gratings, the laser fluence on the final compression has to be kept under the half of its LIDT. Then, this LSIM will affect the maximum output pulse energy from the final grating. As a result, the AFGC is proposed as follows.

**Principle of asymmetric four-grating compressor**
For lasers with low energy and small beam size, a Treacy double-grating compressor is enough for pulse compression [19], where the two gratings parallel to each other, and the incident beam diffracts twice on different positions of each grating with the help of a retroreflector. This double-grating compressor is simple and easy to run as there are only two control varies: the incident angle and the distance between the two gratings. Obviously, no spatial dispersion will be introduced in the output beam owing to the automatically equivalent symmetric FGC configuration.

While for high peak power lasers with large beam size, symmetric TFGCs are usually used in almost all previous PW lasers to make sure output pulses with no spatial dispersion. Generally, a TFGC contains two pairs of gratings and owns a bilateral symmetry structure. Gratings parallel to each other in each pair, and the distances between gratings in each pair are needed to keep the same.

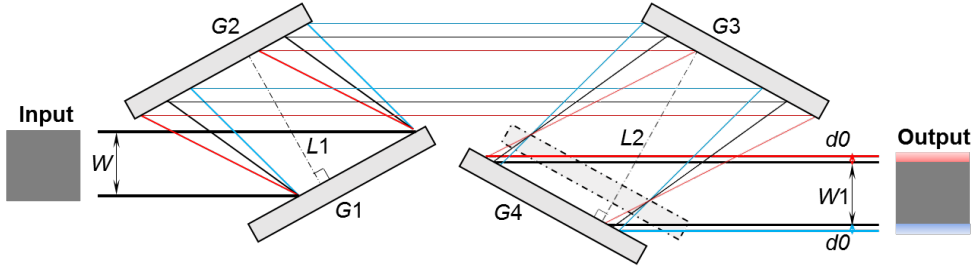

Fig.1, The optical configuration of a FGC. *W*, *W*1, widths with full spectral components along the horizontal direction in the input and output beams, respectively. *G*1-4, diffraction gratings. *L*1, *L*2, perpendicular distances of the first and second pair of gratings, respectively. *d*0, width with partial spectral components along the horizontal direction in the output beam. When *L*1=*L*2, the compressor is a general TFGC, grating with dot dash outline indicates the position of the final grating. When *L*1≠*L*2, the compressor is the proposed AFGC with asymmetric structure, just the condition of *L*1<*L*2 is shown here with *L*2 is increased to make the final grating at the position of *G*4 and *L*1 is decreased simultaneously while the new position of the first grating is not shown here.

The optical configuration of the proposed AFGC is presented in Fig. 1. *L*1 is the distance between the first grating *G*1 and the second grating *G*2, and *L*2 is the distance between the third grating *G*3 and the fourth grating *G*4. *L*1≠*L*2 makes it different from a general TFGC with *L*1=*L*2. The asymmetric structure make the pulses output from the compressor contain spatial dispersion. Spatial dispersion will induce beam smoothing to the output laser beam. And as a result, hot spots or local strong points are smoothed and weakened on the final grating. Of course, to achieve the same spectral dispersive compensation, the sum of *L*1 and *L*2 keeps the same value for both TFGC and AFGC. Currently, most of the pulse compressors in PW lasers are running with general TFGC configuration, of which the input laser beam show no spatial dispersion. Then, we will discuss in detail the condition of AFGC with *L*1<*L*2 here, which can be modified easily based on the widely used general TFGC.

The amount of spatial dispersion *d* of two frequency components $\omega$ and $\omega_0$ introduced by a pair of diffraction gratings is shown in Fig. 2, which can be described as:

$$d = (\tan\beta(\omega_0) - \tan\beta(\omega))\cos\alpha * L. \qquad (1)$$

Then, the induced spatial dispersion *d*0 on one side of a FGC, which is shown in Fig. 1, can be expressed as:

$$d0 = (\tan\beta(\omega_s) - \tan\beta(\omega_l))\cos\alpha * (L2 - L1). \qquad (2)$$

Where $\omega_s$ and $\omega_l$ represent the shortest and the longest wavelength of the input laser beam, $\alpha$ is the incident angle on the grating. For a general symmetric TFGC with *L*1=*L*2, there is no spatial dispersion with *d*0=0, while for an AFGC, *d*0≠0. From the expression, the spatial dispersion is linearly related to the difference between the two perpendicular distances *L*1 and *L*2. However, the increasing of *L*2 will accompanied by the decreasing of *L*1, which means the first grating may induce light obstructing for a small *L*1 that should be carefully considered simultaneously. From the recent work of the MPC method [18], the larger the induced spatial dispersion *d*0, the smoothing or the lower the LSIM of the output laser beam. The same as prism pair, the LSIM is reduced rapidly at first from 2.0 to about 1.3, it needs relative large *d*0 to reduce LSIM from 1.3 to 1.1. Note that large *d*0 means a large size of final grating and increasing of pulse duration after the spatiotemporal focusing at the focal plane. Since the peak power improvement and running safety are the two main advantages of this AFGC method. Then, the *d0* can be set to about 10 mm that is enough to smooth the accident hot spots which usually are about 1 mm in diameter in the final grating. This running safety is very important for most current running PW lasers with TFGC. To improve the output peak power, the *d0* should be smoothed to a relative lower LSIM, where *d0*=60 mm will reduce the LSIM from 2.0 to about 1.1.

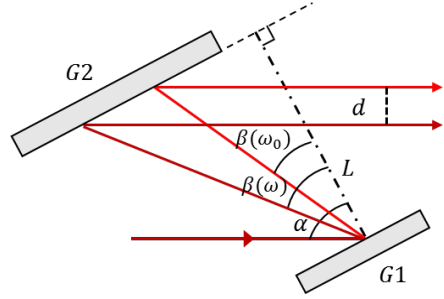

Fig.2. Spatial dispersion introduction by a pair of gratings. G1-2, diffraction gratings. $\alpha$, incident angle. $\beta(\omega)$ and $\beta(\omega_0)$, diffraction angles of the frequencies $\omega$ and $\omega_0$, respectively. L, perpendicular length of G1 and G2. d, amount of spatial dispersion.

We assume the LSIM of the output laser beam can be weakened from about *m*=2 to *m*=1.1, let the maximum laser fluence the final grating can bear to be $F4_{m=2}$ with *m*=2 and $F4_{m=1.1}$ with *m*=1.1, it can be obtained that:

$$2 \times F4_{m=2} = 1.1 \times F4_{m=1.1}. \qquad (3)$$

Considering the laser fluence ratio of the first and final gratings is *F1:F4* = 1.37:1, in the condition of *m*=1.1, the laser fluence on the first grating is:

$$F1 = 1.37 \times \frac{2}{1.1} \times F4_{m=2} \approx 2.5 \times F4_{m=2}. \qquad (4)$$

As the LIDT ratio of the first and final gratings for nanosecond and femtosecond pulses, respectively, is about $F1_{LIDT}: F4_{LIDT}$= 2.67:1 [18], in the condition of *m*=1.1, the maximum laser fluence on the first grating can be:

$$F1_{max} = 2.67 \times F4_{m=2}. \qquad (5)$$

As $F1_{max} > F1$, we can conclude that, if the AFGC introducing spatial dispersion weakens the LSIM from *m*=2 to *m*=1.1, the maximum output energy of the compressor can be increased by about $F4_{m=1.1} / F4_{m=2} \approx 1.8$ times, and the compressor can still work under its LIDT.

**Simulation of the near and far fields of the laser beam**
During the simulation process, an amplified 10$^{th}$ order super-Gaussian profile 500×500 mm² beam with 4 ns chirped pulse duration is used as the input laser into an AFGC. The spectrum of the input laser pulse is 6$^{th}$ order super-Gaussian profile centered at 925 nm with 200 nm full spectral bandwidth, and its full width at half maximum (FWHM) Fourier transform limited (FTL) duration is about 14.5 fs. The compressor consists of four gold-coated gratings with 1400 lines/mm groove density and 90% diffraction efficiency, and the incident angle is 61°. During the simulation, the dimension step is set to be 0.9 mm, which is almost the equal size of hot spots or strong points in real amplified beams. The induced spatial dispersion lengths of *d0*=10 mm and *d0*=60 mm along the horizontal direction (vertical to the grating groove lines) are used for the simulation. According to grating equation and Eq. (2), the distance difference between *L2* and *L1*, that is *L2-L1*, should be about 54 mm and 322 mm for *d0*=10mm and *d0*=60mm, respectively. While for a 4 ns chirped pulse with 200 nm full bandwidth, the total perpendicular length of the two grating pairs, that is *L1+L2*, is about 2480 mm according to the principle of Treacy gratings [19]. Then, the distances *L1* between *G*1 and *G*2 are set to be 1213 mm/1079 mm, and the distances *L2* between *G*3 and *G*4 are set to be 1267 mm/1401 mm as for *d0* is 10mm/60mm, respectively.

Figure 3 shows the near field beam profiles of the input and output beams in the AFGC when the induced spatial dispersion *d0* are about 10 mm and 60 mm, respectively.

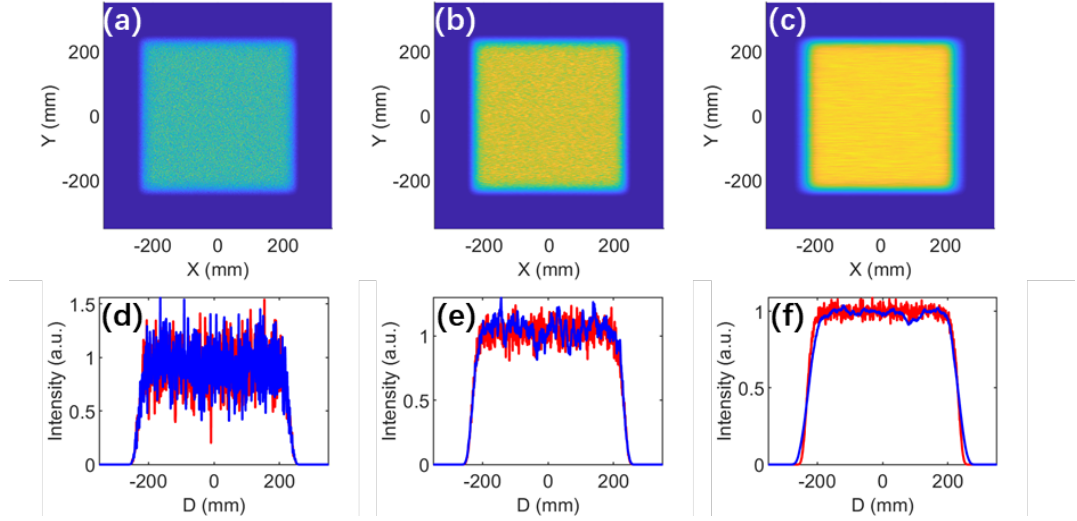

Fig. 3. The near field beam profiles of the input and output pulses of the AFGC. Input beam profile (a), and its intensity curves (d) at spatial positions Y=0 (blue curve) and X=0 (red curve). Output beam profile (b) and its intensity curves (e) at spatial positions Y=0 (blue curve) and X=0 (red curve) at *d0*=10 mm. Output beam profile (c) and its intensity curves (f) at spatial positions Y=0 (blue curve) and X=0 (red curve) at *d0*=60 mm.

The two-dimensional beam profile of the input beam into the AFGC is shown as Fig. 3(a). Figure 3(d) shows its intensity curves at spatial positions Y=0 (blue curve) and X=0 (red curve). As can be seen that the LSIM are about 2.0 on both directions. When *d0* is set to 10 mm, the output two-dimensional beam profile from the AFGC is shown as Fig. 3(b), while Fig. 3(e) shows the corresponding intensity curves of Fig. 3(b) at spatial positions Y=0 (blue curve) and X=0 (red curve). Obviously, the beam is smoothed and the output LSIM is reduced to be about 1.3. In this case, all hot spots will be smoothed which will undoubtedly protect the final grating damage from strong intensity points in all previous PW lasers, where the general TFGC is usually used and can be modified into AFGC very easily.

The two-dimensional beam profile of the laser beam output from the AFGC is shown as Fig. 3(c) when *d0* is set to 60 mm. Figure 3(f) shows the corresponding intensity curves at center position on both directions, that is Y=0 (blue curve) and X=0 (red curve). In this case, the LSIM can be smoothed to be about 1.1. As a result, it can increase much more maximum bearable laser fluence on the final grating, which means a relative higher output pulse energy.

When *d0*=60 mm, the size of the output beam is increased to be about (500+60)×500 mm². From Fig. 3(f), it can be concluded that the equivalent beam size that with full spectral components is about (500+30)×500 mm$^2$. As the femtosecond LIDT laser fluence is about 229 mJ/cm$^2$ for the used gold diffraction grating [18], considering the LSIM of the output beam is about 1.1, the pulse duration is about 15 fs and input angle is 61°, the total output power of the AFGC can then be about (500+30)×500 mm$^2$×229 mJ/cm$^2$ /cos61°/1.1 /15 fs≈70 PW. With a input beam size of 550×700 mm$^2$, the power can be increased to about (550+30)×700 mm$^2$×229 mJ/cm$^2$ /cos61°/1.1 /15 fs≈116 PW.

In the MPC method [18] and the proposed AFGC here, the key point is that the output laser beam after the pulse compressor can own spatial dispersion. It is a conception change that the output laser beam in PW laser do not have to be without spatial dispersion. This is because the induced spatial dispersion of the output beam can be automatically compensated by using the spatiotemporal focusing technique and will be analyzed as follows.

In the near field of the compressed pulse, spatial dispersion introduced by the AFGC will induce spatio-temporal coupling [21], and the temporal (or spectral) properties vary with different positions in the beam for a 500×500 mm$^2$ input laser beam. For a well compressed pulse with a induced spatial dispersion length of $d0$=60 mm, about (500-60)×500 mm$^2$ area in the middle region of the output beam owns the same 200 nm full spectral bandwidth, while for the rest two 60×500 mm$^2$ edge parts, they own continuously narrowed spectrum with spatial dispersion.

Spatiotemporal focusing technique, which has been successfully used in two-photon microscopy and ultrafast micro-machining applications [22, 23], is used for the spatial dispersion compensation of the compressed beam, and it can compensate the spatial dispersion at the focal plane automatically. The compressed output beam is focused by a parabolic mirror with 3 meters focal length to exploring the spatial dispersion influence on the far field of the beam.

A compare of simulated spatiotemporal properties at the focal plane between two beams without and with $d0$=60 mm spatial dispersion is shown in Fig.4. Figure 4(a-c) show spatiotemporal properties at the focal plane when there is spatial dispersion ($d0$=60 mm) introduced along the horizontal direction, and Fig. 4(d-f) show the spatiotemporal properties at the focal plane when there is no spatial dispersion ($d0$=0 mm). It is shown that the introduced spatial dispersion along the horizontal direction induces a slightly pulse front tilt in the *Vertical vs. Time* plane (Fig. 4(b)). In the time domain, the $d0$=60 mm spatial dispersion slightly broaden the pulse duration in the *Horizontal vs. Time* plane as shown in Fig. 4(g), and for the total pulse duration at the focal plane as shown in Fig. 4(i), that can be almost neglected. The red curves in Fig. 4(g-i) show the about 14.5 fs FTL pulse durations without spatial dispersion and the blue curves the pulse durations with spatial dispersion, they almost coincide. In the spatial domain, the focal spots are almost the same for these two beams, as shown in Figure 4(c) and 4(f), with FWHM diameter of 8.5 μm. As a result, both the focal diameter and the pulse duration at the focal plane are barely affected by the induced spatial dispersion. It means that the spatiotemporal focusing technique is a simple but an effective method that can be used to automatically compensate the introduced spatial dispersion of the output compressed beam.

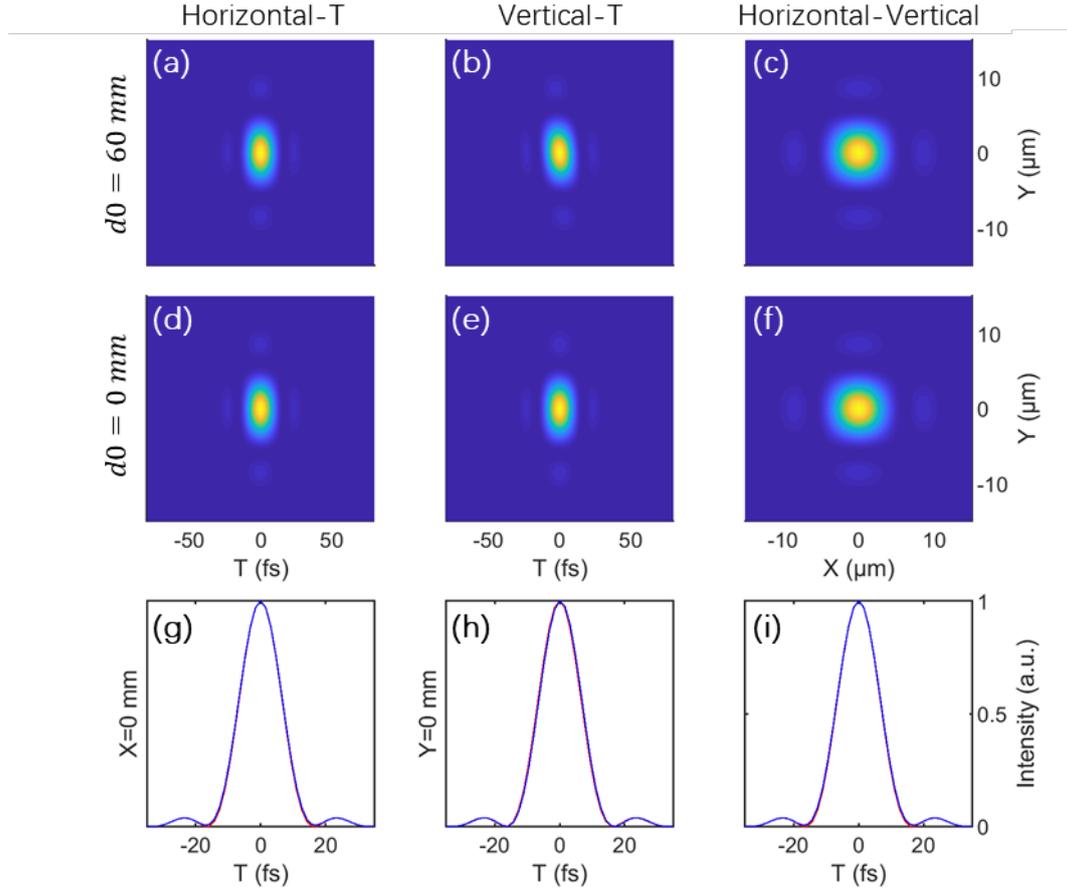

Fig. 4. The spatiotemporal properties at the focal plane with and without spatial dispersion on the laser beam. Spatiotemporal properties, and the integrated spatial profiles at the focal plane in the case of (a-c) with *d0*=60 mm spatial dispersion along the horizontal direction, and (d-f) with no spatial dispersion (*d0*=0 mm), respectively. (g-h): Temporal profiles when (g) X=0 mm in Horizontal-T plane, (h) Y=0 mm in Vertical-T plane, and (i) the integrated temporal profiles at the focal plane. Red curves: without spatial dispersion, blue curves: with *d0*=60 mm spatial dispersion along the horizontal direction.

**Discussion and conclusion**

Laser-induced damage is a key limitation in PW lasers as the strong laser fluence may damage optical components with limited sizes and damage thresholds, where diffraction gratings in pulse compressor have been the shortest stave of PW lasers up to now. This is because manufacturing a high quality meter-sized grating remains particularly challenging [10, 11]. Methods such as tiled grating, CBC and MPC have been studied to break the limitation of diffraction grating manufacture, while companied with extra challenging of control. The proposed AFGC method for PW lasers increases the bearable compressed output pulse energy and running safety with no additional optical component nor extra control than a traditional TFGC.

The method is based on spatiotemporal property modification to decrease LSIM of the laser directly in compressor. According to the simulation section, it is an effective way to decrease the LSIM from about 2 to 1.1 if introduce sufficient spatial dispersion of *d0*=60 mm. And the spatial dispersion introduced can be well compensated at the focal plane by using the spatiotemporal focusing technique. The maximum bearable output energy from the compressor can be increased from two aspects: Firstly, the LSIM decrease from 2 to 1.1 can increase 1.8 times of maximum bearable output energy without of laser-induced damage on the first and final gratings. What's more, compared with a traditional TFGC, the spatial dispersion introduced by the AFGC will increases the beam size on the final grating which results in higher bearable output energy with a modified input beam profile. Based on the widely used traditional TFGC, the AFGC neither requires additional optical components nor increases control difficulty, which makes it an ultrahigh cost performance technique.

In conclusion, an AFGC method that can increase the running safety and the maximum bearable output pulse energy of PW lasers is proposed. The AFGC introduces spatial dispersion in the output beam to decrease the LSIM, which is a key factor that affects the output energy of the compressor. The introduced spatial dispersion can be automatically compensated at the focal plane by using the spatiotemporal focusing technique. With input beam size of 550×700 mm$^2$, 100 PW output power can be achieved in theory. The AFGC method is of ultrahigh cost performance which makes it can be used in almost all current PW lasers. Together with post compression with thin films or thin glass plates [24-26] after AFGC with smoothed laser beam, higher peak power is expected to be obtained in the future.

**Funding**

This work is supported by the National Natural Science Foundation of China (NSFC) (61527821, 61905257, U1930115), and the Strategic Priority Research Program (XDB16) of the Chinese Academy of Sciences (CAS), and Shanghai Municipal Science and Technology Major Project (2017SHZDZX02).

**Disclosures**

The authors declare no conflicts of interest.

**References**

1. D. Strickland and G. Mourou, "Compression of amplified chirped optical pulses," Opt. Commun. **56**, 219-221 (1985).
2. A. Dubietis, G. Jonusauskas, and A. Piskarskas, "Powerful femtosecond pulse generation by chirped and stretched pulse parametric amplification in bbo crystal," Opt. Commun. **88**, 437-440 (1992).
3. C. N. Danson, C. Haefner, J. Bromage, T. Butcher, J.-C. F. Chanteloup, E. A. Chowdhury, A. Galvanauskas, L. A. Gizzi, J. Hein, D. I. Hillier, N. W. Hopps, Y. Kato, E. A. Khazanov, R. Kodama, G. Korn, R. Li, Y. Li, J. Limpert, J. Ma, C. H. Nam, D. Neely, D. Papadopoulos, R. R. Penman, L. Qian, J. J. Rocca, A. A. Shaykin, C. W. Siders, C. Spindloe, S. Szatmari, R. M. G. M. Trines, J. Zhu, P. Zhu, and J. D. Zuegel, "Petawatt and exawatt class lasers worldwide," High Power Laser Sci. Eng. **7**(2019).
4. M. D. Perry, D. Pennington, B. C. Stuart, G. Tietbohl, J. A. Britten, C. Brown, S. Herman, B. Golick, M. Kartz, J. Miller, H. T. Powell, M. Vergino, and V. Yanovsky, "Petawatt laser pulses," Opt. Lett. **24**, 160-162 (1999).
5. W. Li, Z. Gan, L. Yu, C. Wang, Y. Liu, Z. Guo, L. Xu, M. Xu, Y. Hang, Y. Xu, J. Wang, P. Huang, H. Cao, B. Yao, X. Zhang, L. Chen, Y. Tang, S. Li, X. Liu, S. Li, M. He, D. Yin, X. Liang, Y. Leng, R. Li, and Z. Xu, "339 J high-energy Ti:sapphire chirped-pulse amplifier for 10 PW laser

6. F. Lureau, G. Matras, O. Chalus, C. Derycke, T. Morbieu, C. Radier, O. Casagrande, S. Laux, S. Ricaud, G. Rey, A. Pellegrina, C. Richard, L. Boudjemaa, C. Simon-Boisson, A. Baleanu, R. Banici, A. Gradinariu, C. Caldararu, B. D. Boisdeffre, P. Ghenuche, A. Naziru, G. Kolliopoulos, L. Neagu, R. Dabu, I. Dancus, and D. Ursescu, "High-energy hybrid femtosecond laser system demonstrating 2 x 10 PW capability," High Power Laser Sci. Eng. **8**(2020).
7. G. A. Mourou, T. Tajima, and S. V. Bulanov, "Optics in the relativistic regime," Rev. Mod. Phys. **78**, 309-371 (2006).
8. E. Cartlidge, "The light fantastic," Science **359**, 382-385 (2018).
9. B. Shen, Z. Bu, J. Xu, T. Xu, L. Ji, R. Li, and Z. Xu, "Exploring vacuum birefringence based on a 100 PW laser and an x-ray free electron laser beam," Plasma Phys. Controlled Fusion **60**(2018).
10. N. Bonod and J. Neauport, "Diffraction gratings: from principles to applications in high-intensity lasers," Adv. Opt. Photonics **8**, 156-199 (2016).
11. J. Liu, X. Shen, Z. Si, C. Wang, C. Zhao, X. Liang, Y. Leng, and R. Li, "In-house beam-splitting pulse compressor for high-energy petawatt lasers," Opt. Express **28**, 22978-22991 (2020).
12. J. Qiao, A. Kalb, T. Nguyen, J. Bunkenburg, D. Canning, and J. H. Kelly, "Demonstration of large-aperture tiled-grating compressors for high-energy, petawatt-class, chirped-pulse amplification systems," Opt. Lett. **33**, 1684-1686 (2008).
13. N. Blanchot, G. Marre, J. Neauport, E. Sibe, C. Rouyer, S. Montant, A. Cotel, C. Le Blanc, and C. Sauteret, "Synthetic aperture compression scheme for a multipetawatt high-energy laser," Appl. Opt. **45**, 6013-6021 (2006).
14. V. E. Leshchenko, V. A. Vasiliev, N. L. Kvashnin, and E. V. Pestryakov, "Coherent combining of relativistic-intensity femtosecond laser pulses," Appl. Phys. B-Lasers Opt. **118**, 511-516 (2015).
15. N. Blanchot, G. Behar, J. C. Chapuis, C. Chappuis, S. Chardavoine, J. F. Charrier, H. Coic, C. Damiens-Dupont, J. Duthu, P. Garcia, J. P. Goossens, F. Granet, C. Grosset-Grange, P. Guerin, B. Hebrard, L. Hilsz, L. Lamaignere, T. Lacombe, E. Lavastre, T. Longhi, J. Luce, F. Macias, M. Mangeant, E. Mazataud, B. Minou, T. Morgaint, S. Noailles, J. Neauport, P. Patelli, E. Perrot-Minnot, C. Present, B. Remy, C. Rouyer, N. Santacreu, M. Sozet, D. Valla, and F. Laniesse, "1.15 PW-850 J compressed beam demonstration using the PETAL facility," Opt. Express **25**, 16957-16970 (2017).
16. C. Peng, X. Liang, R. Liu, W. Li, and R. Li, "High-precision active synchronization control of high-power, tiled-aperture coherent beam combining," Opt. Lett. **42**, 3960-3963 (2017).
17. D. Wang and Y. Leng, "Simulating a four-channel coherent beam combination system for femtosecond multi-petawatt lasers," Opt. Express **27**, 36137-36153 (2019).
18. J. Liu, X. Shen, S. Du, and R. Li, "A multistep pulse compressor for 10s to 100s PW lasers," arXiv:2103.00843 [physics.optics] (2021).
19. E. B. Treacy, "Optical pulse compression with diffraction gratings," IEEE J. Quantum Electron. **QE 5**, 454-& (1969).
20. P. Poole, S. Trendafilov, G. Shvets, D. Smith, and E. Chowdhury, "Femtosecond laser damage threshold of pulse compression gratings for petawatt scale laser systems," Opt. Express **21**, 26341-26351 (2013).
21. C. Dorrer, "Spatiotemporal Metrology of Broadband Optical Pulses," IEEE J. Sel. Top. Quantum Electron. **25**(2019).

22. F. He, B. Zeng, W. Chu, J. Ni, K. Sugioka, Y. Cheng, and C. G. Durfee, "Characterization and control of peak intensity distribution at the focus of a spatiotemporally focused femtosecond laser beam," Opt. Express **22**, 9734-9748 (2014).
23. G. H. Zhu, J. van Howe, M. Durst, W. Zipfel, and C. Xu, "Simultaneous spatial and temporal focusing of femtosecond pulses," Opt. Express **13**, 2153-2159 (2005).
24. J. Liu, X. Chen, J. Liu, Y. Zhu, Y. Leng, J. Dai, R. Li, and Z. Xu, "Spectrum reshaping and pulse self-compression in normally dispersive media with negatively chirped femtosecond pulses, " Opt. Express14(2), 979-987 (2006).
25. V. Ginzburg, I. Yakovlev, A. Zuev, A. Korobeynikova, A. Kochetkov, A. Kuzmin, M. Sergey, A. Andrey, S. Ilya, K. Efim, and G. Mourou, "Fivefold compression of 250-TW laser pulses, " Phys. Rev. A 101, 013829 (2020).
26. Z. Li, K. Yoshiaki, and J. Kawanaka, "Simulating an ultra-broadband concept for Exawatt-class lasers, " Sci. Rep. 11, 1-16 (2021).